\documentclass[preprint,preprintnumbers,amsmath,amssymb,12pt]{revtex4}
\usepackage{amsfonts}
\usepackage{amssymb}
\usepackage{latexsym}
\usepackage{graphicx}
\usepackage{verbatim}
\usepackage{rotating}
\usepackage{multirow}
\usepackage{booktabs}
\usepackage{blindtext}
\usepackage[colorlinks,allcolors=blue]{hyperref}
\usepackage[usenames,dvipsnames]{color}
\usepackage[active]{srcltx}
\newcommand{\re}[1]{(\ref{#1})}

\begin{document}
\title{HF, DF, TF: Approximating potential curves, calculating rovibrational states}
\author{Laura E. Angeles Gantes}
\email{lauraE\_AG@gmail.com}
\author{Horacio Olivares-Pil\'on}
\email{horop@xanum.uam.mx}
\affiliation{Departamento de F\'isica, Universidad Aut\'onoma Metropolitana-Iztapalapa,
Apartado Postal 55-534, 09340 M\'exico, D.F., Mexico}

\begin{abstract}
An analytical expression for the potential energy curve for the ground state
$X^1\Sigma^+$ of the hydrogen fluoride  molecule (HF) obtained in the framework
of the Born-Oppenheimer approximation is proposed. This analytical expression for
the potential energy curve is based in the two point Pad\'e approximation, which 
correctly reproduces the asymptotic behavior at small $R\rightarrow 0$ and large 
$R\rightarrow\infty$ internuclear distances, and position and depth of potential well,
leading to the accuracy of 4-5 s.d. when compared with experimental data.
The rovibrational spectra of the diatomic molecule HF is calculated by solving the
Schr\"odinger equation for nuclear motion using the Lagrange-mesh method. The ground
state $X^1\Sigma^+$ contains 21 vibrational  states ($\nu,0$)  and 724 rovibrational  
states $(\nu,L)$ with maximal angular momentum equal to 55. The change of reduced mass in the
Schr\"odinger equation for nuclear motion allows us to obtain the rovibrational spectrum of
the ground state potential curve of the deuterium fluoride (DF) and tritium fluoride (TF)
molecules, it contains 1377 and 1967 rovibrational states ($\nu,L$) with maximal angular
momenta 76 and 92, respectively.  Entire rovibrational spectra is presented for the HF 
molecule and its two isotopologues DF and TF.
\end{abstract}
\maketitle
\section{INTRODUCTION}

In the analysis of molecular systems within the Born-Oppenheimer approximation,
the parametric dependence of the total energy  on the nuclear positions defines the
potential energy surface (PES). In practice, the PES  is given at a set of  discrete points in the
space of nuclear  coordinates which are obtained by solving the electronic motion
in the field produced by the fixed nuclei (charged centers).

In the simplest case of  diatomic molecules, the potential energy surface becomes a
potential energy curve (PEC) as a function of the internuclear  distance $R$.
Nowadays,  the development of numerical techniques for quantum-mechanical
calculations,  has made it possible to obtain highly accurate values of the PEC, however,
as is well known, as  the number of electrons increases, the accuracy deteriorates while
the computational time increases dramatically.  An alternative is to extract the PEC from experimental
results using the Rydberg-Klein-Rees (RKR) method~\cite{RR:1932,OK:1932,ALG:1947},
from where the turning points of the  vibrational levels are obtained.
In order to construct a continuous representation of the PEC, the standard approach
consists of fitting the discrete points for PEC by  traditional potentials
($e.g.$ Morse, Lennard-Jones, P\"oschl-Teller,  etc.) or to consider more sophisticated expressions
for potentials~\cite{IGK:2006}  mainly, near equilibrium and sometimes involving the
domain of large distances.

In this work we consider  the ground state $X^1\Sigma^+$ of the Hydrogen Fluoride (HF),
a diatomic molecule composed of a
Hydrogen atom ($1e, 1p$) and a Fluorine atom ($9e, 9p$). This molecule, with $10$ electrons
in total, has been extensively studied both experimentally and theoretically (see for
example~\cite{LD:1973,CH:1990,ZSCH:1991,CH:2006,LP:2002,CH:2015} and references therein).
At present,  the high accuracy achieved experimentally has made it possible to obtain very
accurate potential energy curves.  Theoretically, it has been a challenge to obtain a complete
description from {\it ab-initio} calculations. The average absolute deviation between experimental
and theoretical vibrational energy levels is
$\sim 80$~cm$^{-1}$ ($\sim 4\,\times10^{-4}$~Hartree)~\cite{LP:2002}.

The goal of this paper, is to present an analytical expression for the ground state $X^1\Sigma^+$
of the potential energy curve $V(R)$ in all domain
 of internuclear distances $R$ with a correct asymptotic behavior at small ($R\rightarrow 0$) and
 large ($R\rightarrow\infty$) values of the distance between nuclei.  The analitical expression
$V(R)$ is given in terms of  a two points Pad\'e approximant.  The methodology followed here,
 has been  applied successfully to some specific diatomic molecules: H$_2^+$, H$_2$, HeH and
 LiH~\cite{OT:2017,TO:2019}. With the explicit analytical expression of the PEC, the Schr\"odinger
 equation for nuclear motion (see below) is solved in Lagrange-mesh method  to obtain the
 rovibrational spectrum for this  diatomic molecule.

Making a simple modification in the nuclear Schr\"odinger equation - it occurs due to the fact that we
make studies in the Born-Oppenheimer approximation - the rovibrational spectra $E_{(\nu,L)}$ is also
obtained for the isotopological species with Deuterium (D) and Tritium (T) content:
Deuterium Fluoride (DF) and Tritium Fluoride (TF) molecules, respectively.

\section{The ground state $X^1\Sigma^+$ of the HF molecule}

Let us start  by describing  the two asymptotic limits,  small ($R\rightarrow 0$) and
large ($R\rightarrow\infty$)  internuclear distances
of the potential energy curve for the ground state  $X^1\Sigma^+$
of the heteronuclear diatomic molecule HF.
In the united atom limit ($R\rightarrow 0$),  we have a Neon  atom (Ne) and the dissociation
energy $\tilde{E}$ for small internuclear distances  is given by
\begin{equation}
    \widetilde E=\frac{9}{R}-\varepsilon_0 + O(R^2)\,,
    \label{eq:short_distances_R}
\end{equation}
where the first term is  the repulsive Coulomb interaction between nuclei  and $\varepsilon_0$
is the difference between the energy of the system in the united atom limit (Neon atom $E_{Ne}$)
and the sum of  the energies for each individual components of the molecule, the Hydrogen  ($E_{H}$)
and Fluorine atom ($E_{F}$)
\begin{equation}
\varepsilon_0=E_{\rm Ne} - (E_{H} + E_{F})\,,
\end{equation}
with $E_{\text H}=-0.5$ a.u. and $E_{\text F}=-99.733424$ a.u.~\cite{HKKT:2012}.
On the other hand at large internuclear distances $R\rightarrow\infty$, the molecule dissociates
into a  Hydrogen atom plus a Fluorine atom, and the dissociation energy is given by
\begin{equation}
    \widetilde E=-\frac{C_6}{R^6}+O\left(\frac{1}{R^8}\right),
    \label{eq:large_distances_R}
\end{equation}
where $C_6=7.766$~\cite{ZSCH:1991} is the  \textit{Van der Waals} coefficient.

In order to construct an analytic representation  for the potential energy curve interpolating the
two asymptotic limits $R\rightarrow 0$~\re{eq:short_distances_R} and
$R\rightarrow \infty$~\re{eq:large_distances_R} a two point Pad\'e type approximant
of the form $E(R)=\frac{1}{R}\text{Pade}\left[N/N+5\right]$ is used.
Choosing $N=5$, the explicit expression for $E(R)$ is given by
\begin{eqnarray}
E(R)_{\{3,2\}}&=&\frac{1}{R}\text{Pade}\left[5/10\right](R)\,,  \\
       &=&\frac{9+a_1R+a_2R^2+a_3R^3+a_4R^4-C_6R^5}{R(1+\alpha_1 R+\alpha_2R^2+b_3R^3+b_4R^4+b_5R^5+b_6R^6+b_7R^7+b_8R^8+\alpha_3R^9+R^{10})}\,,\nonumber
\label{eq:Pade}
\end{eqnarray}
where the three constrains $\alpha_i$ ($i=1,2,3$)
\begin{eqnarray}
\alpha_1&=&\frac{a_1-\varepsilon_0}{9}\,,\\
\alpha_2&=&\frac{-a_1\varepsilon_0+\varepsilon_0^2+9a_2}{81}\,,\nonumber\\
\alpha_3&=&-\frac{a_4}{C_6}\,,\nonumber
\end{eqnarray}
guarantee the expected behavior in the series expansion:
the three  terms $R^{-1}$, $R^{0}$ and $R^{1}$ for small internuclear distances
$R\rightarrow 0$~\re{eq:short_distances_R}
and the two terms $R^{-6}$ and $R^{-7}$ for large internuclear distances
$R\rightarrow \infty$~\re{eq:large_distances_R} are reproduced.
The then  free parameters have been obtained through a fit
of the experimental energy values reported in \cite{CH:2006}
with the analytic expression~\re{eq:Pade}.  Their values result in:
\begin{align*}
 a_1 &=2761.87, & a_2&=-2269.79, & a_3&=-193.29, & a_4&=107.967\\
 b_3&=-660.63, & b_4&=1909.71, & b_5&=-1658.58, & b_6&=988.485, \\
 b_7&=-379.066, & b_8&=94.677.
    \label{eq:a_i,b_i}
\end{align*}

\begin{figure}[!b]
	\centering
	\includegraphics[scale=0.3]{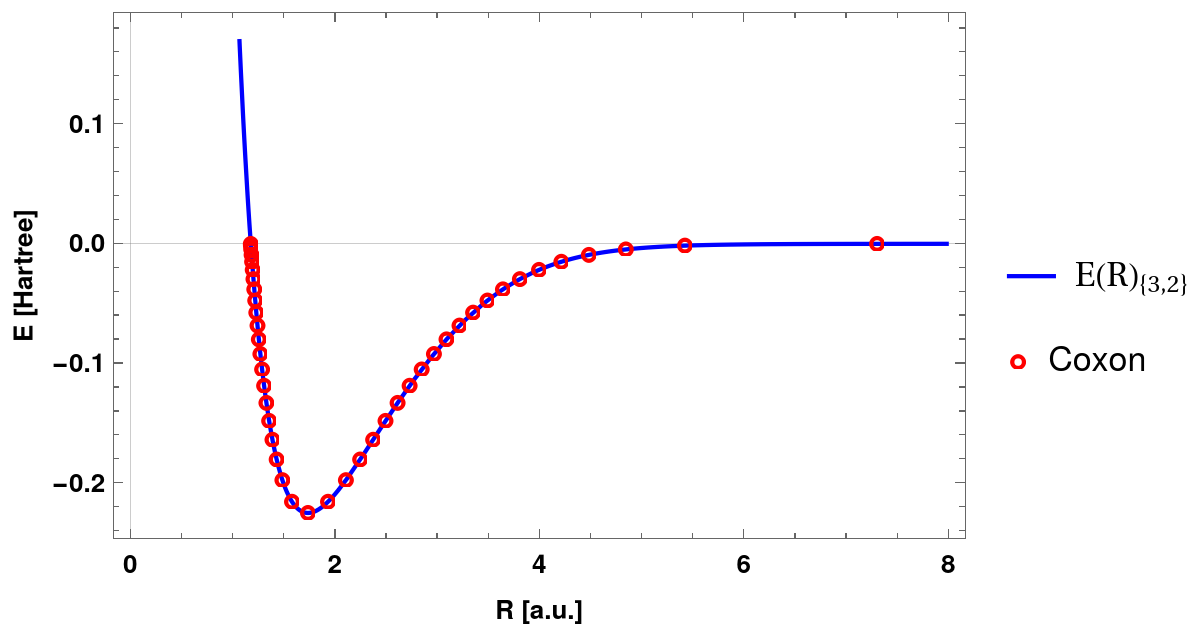}
	\caption{Potential energy curve for the HF molecule.  The solid curve is the fit
	$E(R)_{\{3,2\}}$~\re{eq:Pade},  and the points are the experimental results~\cite{CH:2006}.
	The minimum $E_{min}=-0.224913$~a.u. is located at $R_{eq}=1.73254$~a.u.}
	\label{fig:potential_curve}
\end{figure}

As can be seen in Table \ref{tab:01}, the the analytic expression~\re{eq:Pade} gives
no less that 4 s.d. in correspondence with the experimental results~\cite{CH:2006}.
The position and the deep of the potential energy curve can be calculated by
simply taking the derivative of~\re{eq:Pade} and setting
it equal to zero. As a result, we obtain $R_{eq}=1.73254$~a.u. and $E_{min}=-0.224913$ a.u.
In comparison, the best
numerical result is   $R_{eq}=1.73257$ a.u. and $E_{min}=-0.224908$ a.u.~\cite{CH:2006}.
The analytic expression for the potential energy curve~\re{eq:Pade} as well as the experimental
results \cite{CH:2006}, are depicted in Figure~\ref{fig:potential_curve}.

\begin{table}[t]
  \centering\footnotesize\sffamily
  \caption{Experimental energy of the diatomic molecule HF~\cite{CH:2006} (column 3)
  for their corresponding turning points $R_{\text{min}}$ and $R_{\text{max}}$ of the potential
  energy curve.  Columns 4 and 5 result from fit $E_{\{3,2\}}$~\re{eq:Pade} at
  $R_{\text{min}}$ and $R_{\text{max}}$.}
\scalebox{0.9}{\begin{tabular}{r r r r r}
\hline \multirow{2}{2.5cm}{\centering $R_{\text{min}}$ [u.a.]} &\multirow{2}{2.5cm}{\centering $R_{\text{max}}$ [u.a.]}
 & \multicolumn{3}{p{7.8cm}}%
{\centering E [Hartree] }\tabularnewline \cline{3-5}
& & \multicolumn{1}{p{2.5cm}}%
{\centering Data \cite{CH:2006}}
 & \multicolumn{1}{p{2.5cm}}%
{\centering $E_{\{3,2\}}(R_{\text{min}})$}
& \multicolumn{1}{p{2.5cm}}%
{\centering $E_{\{3,2\}}(R_{\text{max}})$ }
\tabularnewline \hline
1.57634 & 1.92856 & -0.21556 & -0.21557 & -0.21556 \\
1.48248 & 2.10342 & -0.19751 & -0.19752 & -0.19751 \\
1.42634 & 2.24290 & -0.18025 & -0.18025 & -0.18025 \\
1.38529 & 2.36979 & -0.16375 & -0.16374 & -0.16375 \\
1.35286 & 2.49078 & -0.14799 & -0.14798 & -0.14799 \\
1.32613 & 2.60905 & -0.13295 & -0.13295 & -0.13296 \\
1.30352 & 2.72653 & -0.11863 & -0.11863 & -0.11864 \\
1.28405 & 2.84461 & -0.10501 & -0.10502 & -0.10502 \\
1.26707 & 2.96450 & -0.09210 & -0.09210 & -0.09210 \\
1.25212 & 3.08735 & -0.07987 & -0.07987 & -0.07987 \\
1.23890 & 3.21446 & -0.06834 & -0.06835 & -0.06834 \\
1.22716 & 3.34736 & -0.05752 & -0.05752 & -0.05752 \\
1.21673 & 3.48807 & -0.04742 & -0.04742 & -0.04742 \\
1.20748 & 3.63937 & -0.03807 & -0.03807 & -0.03808 \\
1.19934 & 3.80533 & -0.02951 & -0.02951 & -0.02952 \\
1.19225 & 3.99227 & -0.02181 & -0.02181 & -0.02181 \\
1.18618 & 4.21070 & -0.01502 & -0.01502 & -0.01502 \\
1.18116 & 4.47992 & -0.00928 & -0.00927 & -0.00926 \\
1.17725 & 4.84130 & -0.00471 & -0.00470 & -0.00471 \\
1.17457 & 5.41923 & -0.00153 & -0.00153 & -0.00155 \\
1.17337 & 7.29868 & -0.00010 & -0.00010 & -0.00005 \\
\hline
\end{tabular}}
\label{tab:01}
\end{table}

\subsubsection{Rovibrational spectra}
In the Born-Oppenheimer approximation, the rovibrational spectra $E_{(\nu,L)}$ can be obtained
by solving the nuclear Schr\"odinger  equation for  the Hamiltonian
\begin{equation}
   -\frac{1}{2\mu}P^2+\frac{L(L+1)}{2\mu R^2}+V(R)\,,
    \label{eq:Hamiltonian}
\end{equation}
where $L$ is the angular momentum, $\mu$ is the reduced mass and the momentum $P=-i\nabla_R$.
For the concrete case of the Hydrogen Fluoride molecule, the interaction potential $V(R)$ is given by the
analytic expression~\re{eq:Pade} and the reduced mass $\mu=m_H\, m_F/( m_H+ m_F)$
is calculated from the nuclear masses of the Hydrogen and Fluorine atoms:
$m_H=1836.1527$ a.u.  and $m_F=34622.974$ a.u.~\cite{GAT:2003}, respectively.

\begin{figure}[!b]
	\centering
	\includegraphics[scale=1.8]{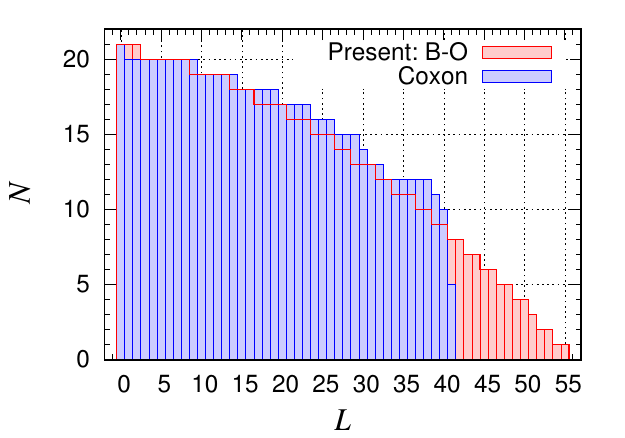}
	\caption{Number of rovibrational states for the ground state X$^1\Sigma^+$ of the 
             HF molecule as a function of the angular momentum $L$ (marked by blue and red).
	         The continuous red line indicates the maximum number of rovibrational states
             obtained by solving the nuclear Schr\"odinger equation.
	         Results from \cite{CH:2015} for $L=0,\dots ,41$, where corrections to the B-O
             approximation are taken into account, are also presented (marked by blue).}
	\label{fig:rovibrational_states}
\end{figure}

In order to solve the one dimension differential equation for the nuclear motion~\re{eq:Hamiltonian},
the Lagrange-mesh method is used~\cite{DB:2015}.  It is found that the potential energy
curve for the ground state $X^1\Sigma^+$ supports  21 pure vibrational states ($L=0$)
$E_{(\nu,0)}$ with $\nu=0,\cdots,20$ in complete agreement with the
experimental results presented in~\cite{CH:2006}  (see  Table~\ref{tab:1}).
The absolute error in the vibrational spectra is  $\lesssim 4\times10^{-5}$.
For comparison,  the experimental data presented in~\cite{LD:1973} are also shown in
Table~\ref{tab:1}.
Figure~\ref{fig:rovibrational_states} shows the number $N$ of vibrational levels  with the
same rotational quantum number $L$. It can be noticed that the number of  rotational levels
$E_{(0,L)}$ of the ground state $X^1\Sigma^+$ is 56 ($L=0, \dots, 55$). In total the potential
energy curve supports  $724$ rovibrational states.  The data from $L=0,\dots ,41$ reported
in~\cite{CH:2015} are also depicted in Figure~\ref{fig:rovibrational_states}.
The presence of  27 extra states between $L=9$ and $L=40$ is due to the fact
that in~\cite{CH:2015} the breakdown of the Born-Oppenheimer approximation
is taken into account.

\begin{table}[t]
  \centering\footnotesize\sffamily
  \caption{Vibrational energies $E_{(\nu, 0)}$ of the ground state $X^1\Sigma^+$ of the
  diatomic molecule HF.}
 \scalebox{0.9}{\begin{tabular}{r r r r}
\hline \multirow{2}{1cm}{\centering$\nu$}
 & \multicolumn{3}{p{10.3cm}}%
{\centering E [Hartree] }\tabularnewline \cline{2-4}
 & \multicolumn{1}{p{3.5cm}}%
{\centering Data \cite{LD:1973}}
 & \multicolumn{1}{p{3.5cm}}%
{\centering Data \cite{CH:2006}}
& \multicolumn{1}{p{3.5cm}}%
{\centering $E_{(\nu,0)}$}
\tabularnewline \hline
0 & -0.21565 & -0.21556 & -0.21556 \\
1 & -0.19760 & -0.19751 & -0.19751 \\
2 & -0.18033 & -0.18025 & -0.18024 \\
3 & -0.16383 & -0.16375 & -0.16373 \\
4 & -0.14807 & -0.14799 & -0.14797 \\
5 & -0.13304 & -0.13295 & -0.13293 \\
6 & -0.11872 & -0.11863 & -0.11861 \\
7 & -0.10510 & -0.10501 & -0.10499 \\
8 & -0.09218 & -0.09210 & -0.09207 \\
9 & -0.07995 & -0.07987 & -0.07984 \\
10 & -0.06842 & -0.06834 & -0.06831 \\
11 & -0.05760 & -0.05752 & -0.05748 \\
12 & -0.04750 & -0.04742 & -0.04739 \\
13 & -0.03815 & -0.03807 & -0.03804 \\
14 & -0.02960 & -0.02951 & -0.02949 \\
15 & -0.02189 & -0.02181 & -0.02178 \\
16 & -0.01511 & -0.01502 & -0.01500 \\
17 & -0.00936 & -0.00928 & -0.00925 \\
18 & -0.00479 & -0.00471 & -0.00469 \\
19 & -0.00161 & -0.00153 & -0.00152 \\
20 &          & -0.00010 & -0.00007 \\
\hline
\end{tabular}}
\label{tab:1}
\end{table}
\section{The ground state $X^1\Sigma^+$ of the DF molecule}

The formalism developed previously for the HF molecule allows us to extend the
study to the isotopologues DF (Deuterium Fluorine) and TF (Tritium Fluorine).
Let's start by considering the deuterium-containing molecule DF.

Due to the fact that we are working inside of the Born-Oppenheimer approximation,
we are not aware of isotopic corrections: the depth of the minimum is the same for all three systems, HF, DF and TF.
For this reason, in order to obtain the rovibrational spectrum by solving the
Schr\"odinger equation for the nuclear motion (with Hamiltonian~\re{eq:Hamiltonian}),
the analytic expression for the  interaction potential is the same as that for the HF
molecule~\re{eq:Pade}.
The difference appears in the nuclear reduced mass
$\mu=m_{D}m_{F}/(m_{D}+m_{F})$~\re{eq:Hamiltonian}, where
 the Deuteron mass is $m_{D}=3670.4833$ a.u. \cite{GAT:2003}.
The resulting differential equation was solved  using the
Lagrange-mesh method~\cite{DB:2015}.

Column 4 of Table~\ref{tab:2}, presents the 29 vibrational states
$E_{(\nu,0)}$, $\nu=0,\dots,28$ supported by the ground state $X^1\Sigma^+$
of the DF molecule. When comparing with the results obtained from experimental
data~\cite{CH:2015}, it is found that the absolute error is $\lesssim 6\times10^{-5}$.
For comparison purposes, column 2 of Table~\ref{tab:2} display the results
given in~\cite{FVM:1960}.

\begin{figure}[!b]
	\centering
	\includegraphics[scale=1.8]{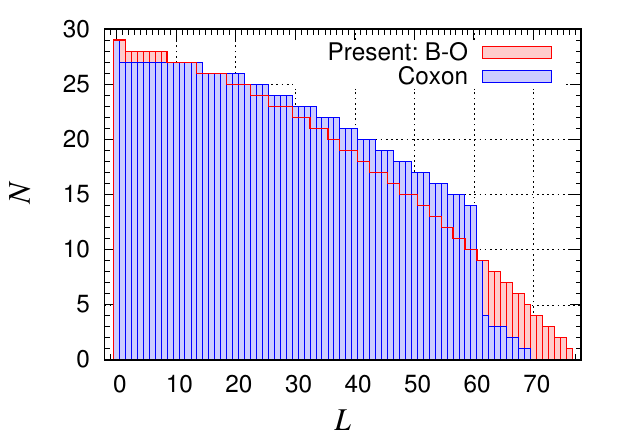}
	\caption{Rovibrational states in the ground state X$^1\Sigma^+$ of the DF molecule
	as a function of the angular momentum $L$.  The continuous red line indicates the maximum
	number of rovibrational states at given $L$ obtained by solving the nuclear Schr\"odinger
     equation. Results reported in~\cite{CH:2015} from $L=0,\dots ,69$, where corrections to the B-O approximation are taken into account, are also presented (marked by blue).  }
	\label{fig:rovibrational_states_DF}
\end{figure}

The number of vibrational states as a function of the angular  number $L$
are depicted in Figure~\ref{fig:rovibrational_states_DF}.
There are 77 pure rotational states $E_{(0,L)}$, with $L=0,\dots,76$.
In total, the ground state $X^1\Sigma^+$  of the DF molecule  supports $1377$
rovibrational states. Corrections to the Born-Oppenheimer approximation taken into account in~\cite{CH:2015} explain the presence of the $87$ additional  states between $L=14$ and $L=60$ shown in Fig.\ref{fig:rovibrational_states_DF}.

\begin{table*}[t]
  	\centering\footnotesize\sffamily
  	\caption{Vibrational energies $E_{(\nu, 0)}$ of the ground state $X^1\Sigma^+$ of the
  	diatomic molecule DF.}
\scalebox{0.85}{\begin{tabular}{r r r r}
\hline \multirow{2}{1.2cm}{\centering$\nu$}
 & \multicolumn{3}{p{10.5cm}}%
{\centering E [Hartree] }\tabularnewline \cline{2-4}
 & \multicolumn{1}{p{3.5cm}}%
{\centering Data \cite{FVM:1960}}
 & \multicolumn{1}{p{3.5cm}}%
{\centering Data~\cite{CH:2015}}
& \multicolumn{1}{p{3.5cm}}%
{\centering $E_{(\nu,0)}$ }
\tabularnewline \hline
0  & -0.21807 & -0.21806 & -0.21812 \\
1  & -0.20483 & -0.20482 & -0.20487 \\
2  & -0.19200 & -0.19199 & -0.19204 \\
3  & -0.17953 & -0.17957 & -0.17962 \\
4  & -0.16744 & -0.16755 & -0.16760 \\
5  & -0.15574 & -0.15593 & -0.15597 \\
6  & -0.14441 & -0.14469 & -0.14473 \\
7  & -0.13346 & -0.13383 & -0.13387 \\
8  & -0.12289 & -0.12335 & -0.12339 \\
9  & -0.11271 & -0.11324 & -0.11328 \\
10 & -0.10290 & -0.10349 & -0.10353 \\
11 & -0.09347 & -0.09411 & -0.09415 \\
12 & -0.08443 & -0.08510 & -0.08513 \\
13 & -0.07576 & -0.07645 & -0.07648 \\
14 & -0.06747 & -0.06817 & -0.06819 \\
15 & -0.05957 & -0.06026 & -0.06028 \\
16 & -0.05255 & -0.05272 & -0.05275 \\
17 & -0.04540 & -0.04558 & -0.04560 \\
18 & -0.03865 & -0.03883 & -0.03885 \\
19 & -0.03231 & -0.03249 & -0.03251 \\
20 & -0.02641 & -0.02659 & -0.02661 \\
21 & -0.02097 & -0.02115 & -0.02117 \\
22 & -0.01603 & -0.01620 & -0.01621 \\
23 & -0.01162 & -0.01178 & -0.01179 \\
24 &          & -0.00794 & -0.00795 \\
25 &          & -0.00475 & -0.00475 \\
26 &          & -0.00228 & -0.00228 \\
27 &          & -0.00066 & -0.00066 \\
28 &          & -0.00006 & -0.00002 \\
\hline
\end{tabular}}
        \label{tab:2}
\end{table*}
\section{The ground state $X^1\Sigma^+$ of the TF molecule}
Let us now consider the diatomic molecule TF which contains  a Tritium atom and a
Fluorine atom. The rovibrational spectra  can be calculated in the frame of the
Lagrange-mesh method~\cite{DB:2015} by solving the
Schr\"odinger equation for the nuclear motion
with Hamiltonian~\re{eq:Hamiltonian} and potential curve~\re{eq:Pade}.
For this molecule, $\mu=m_{T}m_{F}/(m_{T}+m_{F})$, where  the Triton mass is
$m_{T}=  5496.9221$~a.u.

As is shown in Table~\ref{tab:3}, in agreement with the experimental
results~\cite{CH:2015}, the ground state $X^1\Sigma^+$  of the TF molecule
contains 35 pure vibrational states ($L=0$), $E_{(\nu,0)}$ ($\nu=0,\dots34$).
The absolute error between $E_{(\nu,0)}$ and the experimental results is
$\lesssim 8\times10^{-5}$. It is found that there are 93 rotational
states ($\nu=0$), $E_{(0,L)}$ ($L=0,\dots,92$)  and in total the ground state keeps 1967
rovibrational states.

\begin{figure}[!b]
	\centering
	\includegraphics[scale=1.8]{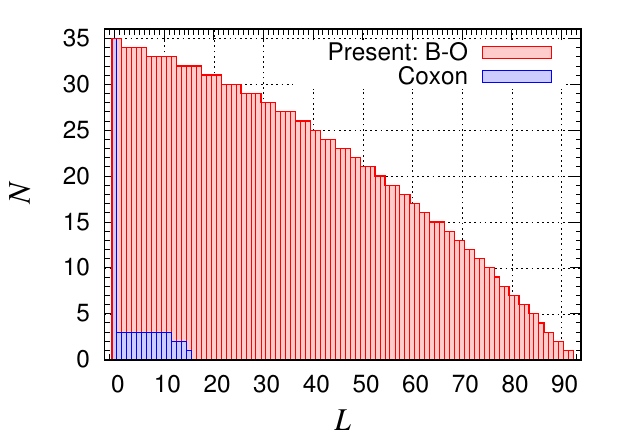}
	\caption{Number of vibrational states for the ground state X$^1\Sigma^+$ of the TF molecule
	as a function of the angular momentum $L$. Results reported in~\cite{CH:2015}
    are also presented (marked by blue). }
	\label{fig:rovibrational_states_TF}
\end{figure}

The number $N$ of vibrational levels as a function of the angular  momentum $L$
is depicted in histogram of Fig.\ref{fig:rovibrational_states_TF} together with those  
reported in~\cite{CH:2015}.

\begin{table*}[t]
  	\centering\footnotesize\sffamily
  	\caption{Vibrational energies $E_{(\nu, 0)}$ of the ground state $X^1\Sigma^+$ of the
  	Tritium Fluorine molecule TF.}
\scalebox{0.78}{\begin{tabular}{r r r}
\hline \multirow{2}{1.2cm}{\centering$\nu$}
 & \multicolumn{2}{p{7.5cm}}%
{\centering E [Hartree] }\tabularnewline \cline{2-3}
 & \multicolumn{1}{p{3.5cm}}%
{\centering Data~\cite{CH:2015}}
& \multicolumn{1}{p{3.5cm}}%
{\centering $E_{(\nu,0)}$ }
\tabularnewline \hline
0  & -0.21915 & -0.21922 \\
1  & -0.20801 & -0.20809 \\
2  & -0.19717 & -0.19724 \\
3  & -0.18662 & -0.18669 \\
4  & -0.17635 & -0.17641 \\
5  & -0.16635 & -0.16642 \\
6  & -0.15664 & -0.15670  \\
7  & -0.14719 & -0.14725 \\
8  & -0.13801 & -0.13807 \\
9  & -0.12909 & -0.12915 \\
10 & -0.12044 & -0.12050  \\
11 & -0.11204 & -0.11210  \\
12 & -0.10390  & -0.10396 \\
13 & -0.09602 & -0.09607 \\
14 & -0.08839 & -0.08844 \\
15 & -0.08102 & -0.08107 \\
16 & -0.07390  & -0.07395 \\
17 & -0.06704 & -0.06709 \\
18 & -0.06044 & -0.06049 \\
19 & -0.05411 & -0.05415 \\
20 & -0.04804 & -0.04808 \\
21 & -0.04225 & -0.04229 \\
22 & 0.03674  & -0.03678 \\
23 & -0.03153 & -0.03157 \\
24 & -0.02662 & -0.02666 \\
25 & -0.02204 & -0.02207 \\
26 & -0.01779 & -0.01782 \\
27 & -0.01391 & -0.01393 \\
28 & -0.01041 & -0.01042 \\
29 & -0.00733 & -0.00734 \\
30 & -0.00471 & -0.00472 \\
31 & -0.00260  & -0.00261 \\
32 & -0.00107 & -0.00107 \\
33 & -0.00024 & -0.00020  \\
34 & -0.00002 & -0.00002  \\
\hline
\end{tabular}}
        \label{tab:3}
\end{table*}

\section{CONCLUSIONS}
Applying the methodology developed in~\cite{OT:2017,TO:2019},  an accurate analytical
representation of the potential energy curve for the ground stated $X^1\Sigma^+$
of the Hydrogen Fluorine  (HF) molecule is presented.
The approximation is based on a two-point P\'ade approximant which satisfies 
the two criteria: 
($i$) it should reproduce correctly the coefficients in the series 
expansion for small ($R \rightarrow 0$) and large ($R\rightarrow \infty$) internuclear distances and 
($ii$) two characteristics of the potential well, the equilibrium position $R_{eq}$ and the dissociation energy $E_{min}$, are reproduced accurately.
Finally, the obtained analytical curve is accurate with not less than 4 s.d. compared 
with experimental results whenever is possible. It is shown that inside of the Born-Oppenheimer approximation the potential energy curve keeps 21 vibrational states ($\nu=0,\dots,20,L=0$), 56 rotational states ($\nu=0,L=0\dots,55$) and, in total, 722 rovibrational states ($\nu,L$).

Modification of the reduced mass $\mu$ in the Schr\"odinger equation for nuclear
motion~\re{eq:Hamiltonian} allows us to explore the isotopologues of the HF molecule.
It is found that the ground state $X^1\Sigma^+$ of the Deuterium Fluorine (DF)/
Tritium Fluorine molecule (TF) supports  29/35 vibrational states $E_{(\nu,0)}$ and
77/93 rotational states $E_{(0,L)}$, respectively.
The number of vibrational states $E_{(\nu,0)}$ calculated for the HF, DF and TF molecules
is in agreement with those reported in~\cite{CH:2006,CH:2015}. On the other hand, the maximum
value of the angular momentum for rotational states $E_{(0,L)}$ is $L_{max}=55, 76$ and $92$,
respectively, (note that in~\cite{CH:2015} the rovibrational states are presented up to
$L = 41, 69$ and $15$) for HF, DF and TF diatomic molecules. In the Born-Oppenheimer
approximation, the total number of rovibrational states supported by the ground state potential
curve are: 724, 1377 and 1967, respectively.
For HF and DF, the difference with the results of~\cite{CH:2015} in the number of vibrational
states for a given angular momentum  $L$
(Figures~\ref{fig:rovibrational_states} and \ref{fig:rovibrational_states_DF}), when a comparison is possible, is due to taken into account the non-adiabatic terms in the Schr\"odinger equation for the nuclear motion.

Finally, it is worth mentioning that the same methodology can be applied to study
the potential energy curves for both, attractive and repulsive excited states of the HF molecule and its isotopologies DF and TF.  Note that with the analytical expressions for the PEC, radiative transitions between different electronic states can be calculated. 
It will be done elsewhere.

\section*{ACKNOWLEDGMENTS}

The authors thank A. Turbiner for bringing the problem to their attention through the work of 
O. Polyansky about the HF dimer, the interest to the
work and for the numerous valuable discussions. HOP also thanks the support through
the Programa Especial de Apoyo a la Investigaci\'on 2019, Universidad Aut\'onoma
Metropolitana. LAG is grateful to CONACyT (Mexico) for a graduate scholarship.

\end{document}